\documentclass[aps,twocolumn,superscriptaddress]{revtex4}
\usepackage{epsfig}
\usepackage{times}
\begin{document}

\title{Biodiversity in models of cyclic dominance is preserved by heterogeneity in site-specific invasion rates}

\author{Attila Szolnoki}
\email{szolnoki@mfa.kfki.hu}
\affiliation{Institute of Technical Physics and Materials Science, Research Centre for Natural Sciences, Hungarian Academy of Sciences, P.O. Box 49, H-1525 Budapest, Hungary}

\author{Matja{\v z} Perc}
\email{matjaz.perc@uni-mb.si}
\affiliation{Faculty of Natural Sciences and Mathematics, University of Maribor, Koro{\v s}ka cesta 160, SI-2000 Maribor, Slovenia}
\affiliation{CAMTP -- Center for Applied Mathematics and Theoretical Physics, University of Maribor, Krekova 2, SI-2000 Maribor, Slovenia}

\begin{abstract}
Global, population-wide oscillations in models of cyclic dominance may result in the collapse of biodiversity due to the accidental extinction of one species in the loop. Previous research has shown that such oscillations can emerge if the interaction network has small-world properties, and more generally, because of long-range interactions among individuals or because of mobility. But although these features are all common in nature, global oscillations are rarely observed in actual biological systems. This begets the question what is the missing ingredient that would prevent local oscillations to synchronize across the population to form global oscillations. Here we show that, although heterogeneous species-specific invasion rates fail to have a noticeable impact on species coexistence, randomness in site-specific invasion rates successfully hinders the emergence of global oscillations and thus preserves biodiversity. Our model takes into account that the environment is often not uniform but rather spatially heterogeneous, which may influence the success of microscopic dynamics locally. This prevents the synchronization of locally emerging oscillations, and ultimately results in a phenomenon where one type of randomness is used to mitigate the adverse effects of other types of randomness in the system.
\end{abstract}

\maketitle

\section*{Introduction}
Models of cyclic dominance are traditionally employed to study biodiversity in biologically inspired settings \cite{frachebourg_prl96, frean_prsb01, kerr_n02, kirkup_n04, mobilia_pre06, reichenbach_n07, mobilia_jtb10, kelsic_n15}. The simplest such model is the rock-paper-scissors game \cite{szolnoki_jrsif14}, where rock crashes scissors, scissors cut paper, and paper wraps rock to close the loop of dominance. The game has no obvious winner and is very simple, yet still, it is an adequate model that captures the essence of many realistic biological systems. Examples include the mating strategy of side-blotched lizards \cite{sinervo_n96}, the overgrowth of marine sessile organisms \cite{burrows_mep98}, genetic regulation in the repressilator \cite{elowitz_n00}, parasitic plant on host plant communities \cite{cameron_jecol09}, and competition in microbial populations \cite{durrett_jtb97, kirkup_n04, neumann_gf_bs10, nahum_pnas11}. Cyclical interactions may also emerge spontaneously in the public goods game with correlated reward and punishment \cite{szolnoki_prx13}, in the ultimatum game \cite{szolnoki_prl12}, and in evolutionary social dilemmas with jokers \cite{requejo_pre12b} or coevolution \cite{szolnoki_epl09, perc_bs10}.

An important result of research involving the rock-paper-scissors game is that the introduction of randomness into the interaction network results in global oscillations \cite{szabo_jpa04, szolnoki_pre04b}, which often leads to the extinction of one species, and thus to the destruction of the closed loop of dominance that sustains biodiversity. More precisely, in a structured population where the interactions among players are determined by a translation invariant lattice, the frequency of every species is practically time-independent because oscillations that emerge locally can not synchronize and come together to form global, population-wide oscillations. However, if shortcuts or long-range interactions are introduced to the lattice, or if the original lattice is simply replaced by a small-world network \cite{watts_dj_n98}, then initially locally occurring oscillations do synchronize, leading to global oscillations and to the accidental extinction of one species in the loop, and thus to loss of biodiversity \cite{ying_cy_jpa07, lutz_jtb13, sun_rs_cpl09, rulquin_pre14, demirel_epjb11}. If the degree distribution of interaction graph is seriously heterogeneous, however, then such kind of heterogeneity can facilitate stable coexistence of competing species \cite{masuda_pre06}. Interestingly, other type of randomness, namely the introduction of mobility of players, also promotes the emergence of global oscillations that jeopardize biodiversity \cite{reichenbach_n07, reichenbach_jtb08, wang_wx_pre11}.

Interestingly, however, although long-range interactions and small-world networks abound in nature, and although mobility is an inherent part to virtually all animal groups, global oscillations are rarely observed in actual biological systems. It is thus warranted to search for universal features in models of cyclic dominance that work in the opposite way of the aforementioned types of randomness. The questions is, what is the missing ingredient that would prevent local oscillations to synchronize across the population to form global oscillations? Preceding research has already provided some possible answers. For instance Peltom{\"a}ki and Alava observed that global oscillations do not occur if the total number of players is conserved \cite{peltomaki_pre08}. Mobility, for example, then has no particular impact on biodiversity because oscillations are damped by the conservation law. However, the consequence of the conservation law does not work anymore if a tiny fraction of links forming the regular lattice is randomly rewired \cite{szolnoki_pre16}. Zealots, on the other hand, have been identified as a viable means to suppress global oscillations in the rock-paper-scissors game in the presence of both mobility and interaction randomness \cite{verma_pre15, szolnoki_pre16}. In addition to these examples, especially in the realm of statistical physics, there is a wealth of studies on the preservation and destruction of biodiversity in models of cyclic dominance \cite{frean_prsb01, de-oliveira_prl02, masuda_jtb08, reichenbach_pre06, peltomaki_pre08b, laird_jtb09, berr_prl09, ni_x_c10, mathiesen_prl11, avelino_pre12, jiang_ll_pla12, roman_jsm12, szczesny_epl13, avelino_pre12b, juul_pre12, laird_e08, roman_pre13, hua_epl13, knebel_prl13, juul_pre13, avelino_pre14, szczesny_pre14, rulquin_pre14, bose_ijb15, javarone_epjb16, knebel_nc16, roman_jtb16}.

Here we wish to extend the scope of this research by considering a partly overlooked property, namely the consideration of site-specific heterogeneous invasion rates. Importantly, we wish to emphasize an important distinction to species-specific heterogeneous invasion rates, which have been considered intensively before. In the latter case, different pairs of species are characterized by different invasion rates, but these differences are then applied uniformly across the population. In case of spatially variable invasion rates, these could be site-specific, and hence particular pairs of species may have different invasion rates even though they are of the same type. Such a setup has many analogies in real life, ranging from differing resources, quality or quantity wise, to variations in the environment, all of which can significantly influence the local success rate of the governing microscopic dynamics. 

Notably, this kind of heterogeneity was already studied in a two-species Lotka-Volterra-like system \cite{dobramysl_prl08}, and in a three-species cyclic dominant system where a lattice has been used as the interaction network \cite{he_q_pre10}. The latter work concluded that the invasion heterogeneity in spatial rock-paper-scissors models has very little effect on the long-time properties of the coexistence state. In this paper, we go beyond the lattice interaction topology, exploring the consequences of quenched and annealed  randomness being present in the interaction network. In the latter case, as we will show, it could be a decisive how heterogeneity is introduced into the invasion rate because annealed randomness does not change the oscillation but quenched heterogeneity can mitigate the global oscillation effectively. In what follows, we first present the main results and discuss the implications of our research, while details concerning the model and the methodology are described in the Methods section.

\section*{Results}

\begin{figure}
\centerline{\epsfig{file=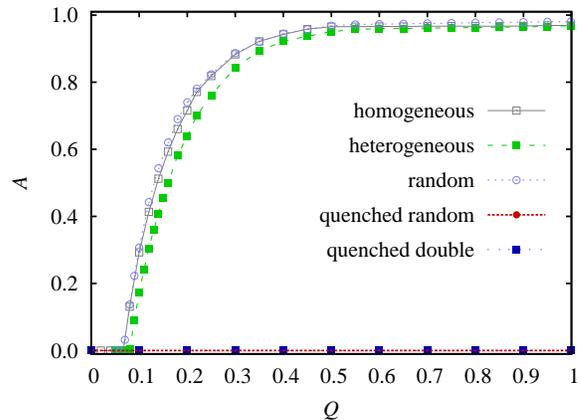,width=8.5cm}}
\caption{\label{suppressed} The emergence of global oscillation due to the introduction of shortcuts to the regular square lattice, as observed in different versions of the rock-paper-scissors game (see legend). Here $Q$ denotes the fraction of rewired links, while $A$ is the area of the limit cycle in the ternary diagram (see Methods). In the homogeneous case all species-specific invasion rates $\delta_i$ are equal to 1, while in the heterogeneous case we use $\delta_0=0.6$, $\delta_1=0.8$, and $\delta_2=1$ throughout the simulation. With random, we denote the case where $\delta_i$ are chosen uniformly at random from the unit interval at each particular instance of the game. Lastly, quenched random and quenched double correspond to results obtained with site-specific invasion rates drawn from a uniform distribution and from a discrete double-peaked distribution, respectively. Importantly, site-specific invasion rates are determined randomly only once at the start of the game and are thereafter quenched. It can be observed that species-specific invasion rates fail to block the emergence of global oscillations, with the order parameter $A$ rising almost to $1$ for a sufficiently high fraction of rewired links. Conversely, site-specific invasion rates completely suppress the emergence of global oscillations, and this regardless of the underlying distribution. As it is specified in the Method section, to produce this plot we have used different system sizes to avoid artificial extinction of species at high oscillation.}
\end{figure}

We first consider results obtained with species-specific invasion rates. Indeed, it is possible to argue that it is too idealistic to assume homogenous invasion rates between different species, and that it would be more realistic to assume that these invasion rates are heterogeneous. But as results presented in Fig.~\ref{suppressed} show, this kind of generalization does not bring about a mechanism that would suppress global oscillations. These oscillations clearly emerge for homogeneous species-specific invasion rates, as soon as the fraction of rewired links of the square lattice $Q$ exceeds a threshold. If we then assume that species-specific invasion rates are heterogeneous, say $\delta_0=0.6$, $\delta_1=0.8$, and $\delta_2=1$ (here $\delta_i$ denotes the invasion rates of $S_i \to S_{i+1}$ transition where $i$ runs from $0$ to $2$ in a cyclic manner), it can be observed that nothing really changes. In fact, the threshold in $Q$ remains much the same, and the order parameter $A$ (the area of the limit cycle in the ternary diagram) reaches the same close to $1$ plateau it does when these invasion rates are homogenous. Further along this line, we can even adopt invasion rates that are chosen uniformly at random from the unit interval at each particular instance of the games. More precisely, we still keep the original $S_i \to S_{i+1}$ direction of invasion, but the strength of the invasion rate $\delta_i$ is chosen randomly in each particular case. But no matter the fact that this rather drastically modifies the microscopic dynamics, the presence of shortcuts will still trigger global oscillations (marked random in Fig.~\ref{suppressed}). We thus arrive at the same conclusion that was already pointed out in \cite{he_q_pre10}, which is that heterogeneous invasion reaction rates have very little effect on the dynamics and the long-time properties of the coexistence state.

\begin{figure*}
\centerline{\epsfig{file=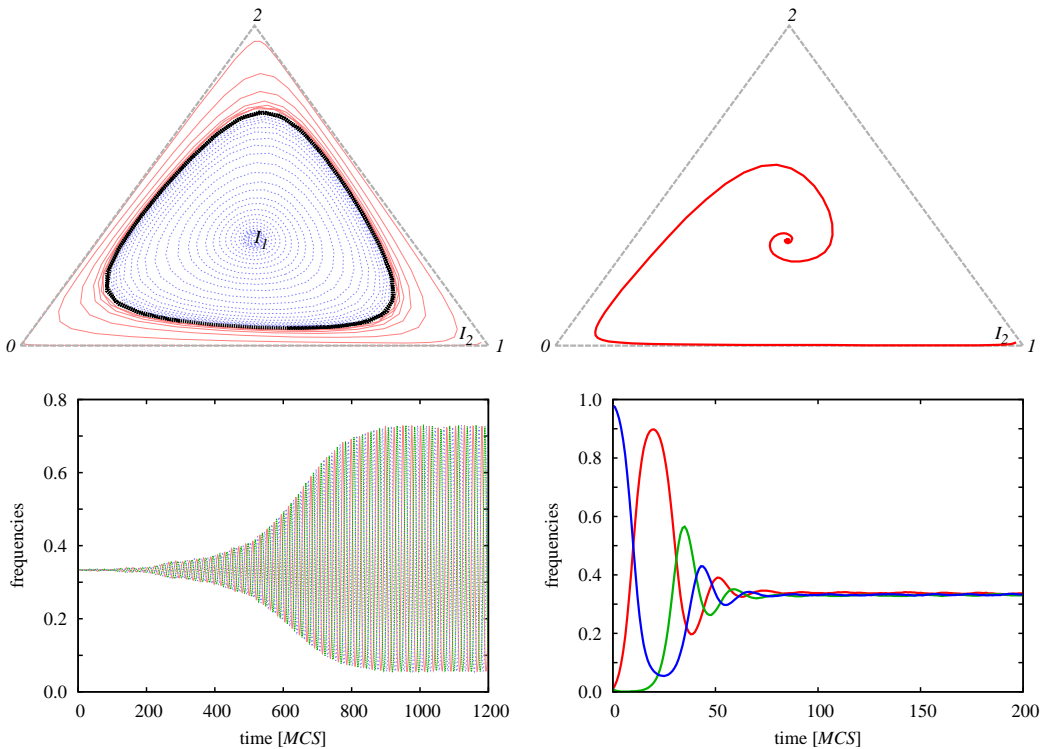,width=14cm}}
\caption{\label{ternary} Evolution of the three species in the ternary diagram (upper row) and time-wise (bottom row), as obtained for a representative example of species-specific invasion rates $\delta_i$ drawn uniformly at random from the unit interval at each particular instance of the game and with site-specific invasion rates $w_j$ drawn uniformly at random from the unit interval once at the start of the game. Note here that if species-specific invasion rates are randomly determined at each particular instance of the game, the distinction to site-specific invasion rates actually is no longer needed since in fact in both cases invasion rates can be considered as site-specific, only that in one case they change over time (left), while in the other case they are quenched (right). All other simulation details are the same in both cases. Namely, a fraction $Q=0.15$ of the links forming the square lattice has been rewired once to yield an interaction network with small-world properties that support the emergence of global oscillations. As can be observed, time-varying interaction rates fail to suppress these oscillations, as the limit cycle and the large-amplitude oscillations are clearly visible on the left. More precisely, the limit cycle, marked black in the ternary diagram on the left, is reached regardless of the initial state ($I_1$ inside the cycle and $I_2$ outside). Both panels of the right show a dramatically different time evolution. In this case the system always terminates into the center of the ternary diagram, and the corresponding time course (bottom right) clearly converges to a steady state. To produce these trajectories we have used a system contains $10^6$ nodes.}
\end{figure*}

Having established the ineffectiveness of heterogeneous species-specific invasion rates to prevent local oscillations to synchronize across the population to form global oscillations, we next consider site-specific heterogeneous interaction rates, denoted as $w_j$ and applied to each site $j$. Here $w_j$ determines the probability that a neighbor will be successful when trying to invade player $j$ according to the original $S_i \to S_{i+1}$ rule. As such, different values of $w_j$ influence the success of microscopic dynamics locally. Moreover, these invasion rates are determined once at the start of the game and can be drawn from different distributions. The simplest case is thus to consider values drawn uniformly at random from the unit interval. As results in Fig.~\ref{suppressed} show (see quenched random), this modification of the rock-paper-scissors game clearly blocks the emergence of global oscillations regardless of the value of $Q$. Indeed, even if the square lattice is, through rewiring, transformed into a regular random graph, the order parameter $A$ still remains zero. Even if the uniform distribution is replaced by a simple discrete double-peaked distribution (practically it means that half of the players has $w_j=0.1$ for example, while the other half retains $w_j=1$), the global oscillations never emerge (see quenched double in Fig.~\ref{suppressed}). The coordination effect leading up to global oscillations is thus very effectively disrupted by heterogeneous site-specific invasion rates, and this regardless of the distribution from which these rates are drawn.

To illustrate the dramatically contrasting consequences of different types of randomness, we show in Fig.~\ref{ternary} representative time evolutions for both cases. The comparison reveals that, as deduced from the values of the order parameter $A$ displayed in Fig.~\ref{suppressed}, time-varying invasion rates fail to suppress global oscillations, the emergence of which is supported by the small-world properties of the interaction network (ternary diagram and the time course on the left). The limit cycle denoted black in the ternary diagram and the large-amplitude oscillations of the densities of species in the corresponding bottom panel clearly attest to this fact. This stationary state is robust and is reached independently of the initial mixture of competing strategies. Conversely, quenched heterogeneous interaction rates drawn from a uniform distribution clearly suppress global oscillations (ternary diagram and the time course on the right). Here, the system will always evolve into the $\rho_i=1/3$ state, central point of the diagram, even if we launched the evolution from a biased initial state. Thus, if heterogeneities are fixed in space, just like in several realistic biological systems, then this effectively prohibits global oscillation by disrupting the organization of a coordinated state, i.e., synchronization of locally occurring oscillations across the population.

As demonstrated previously \cite{szolnoki_njp15, szolnoki_pre16}, the type of randomness in the interaction network responsible for the emergence of global oscillations plays a negligible role. Be it quenched through the one-time rewiring of a fraction $Q$ of links forming the original translation invariant lattice, or be it annealed through the random selection of far-away players to replace nearest neighbors as targets of invasion with probability $P$, there exist a critical threshold in both where global oscillations emerge if invasion rates are homogeneous. Accordingly, it makes sense to test whether heterogeneous site-specific invasions rates are able to suppress such oscillations regardless of the type of randomness that supports them. To that effect, we make use of the discrete double-peaked distribution, where the fraction of sites $\nu$ having a lower invasion rate $w_j=0.1$ then the rest of the population at $w_j=1$ can be a free parameter determining the level of heterogeneity. Evidently, at $\nu=0$ we retain the traditional rock-paper-scissors game with homogeneous invasion rates (all sites have $w_j=1$), while for $\nu>0$ the fraction of sites having $w_j=0.1$, and thus the level of heterogeneity, increases. At the other extreme, for $\nu=1$, we of course again obtain a homogeneous population where everybody has $w_j=0.1$, but we do not explore this option since it is practically identical, albeit the evolutionary process is much slower. By introducing heterogeneity into the system gradually, we can monitor how it influences the stationary state.

\begin{figure}
\centerline{\epsfig{file=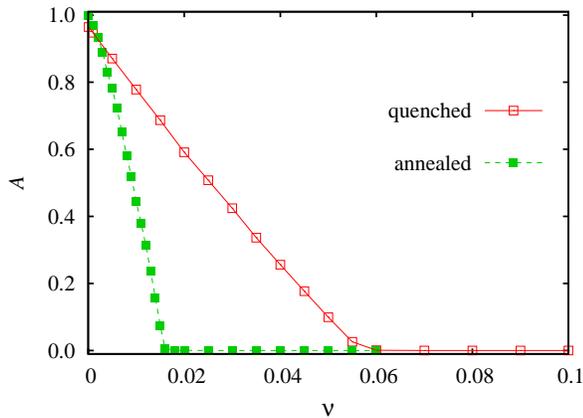,width=8.5cm}}
\caption{\label{nu} The suppression of global oscillations in the presence of quenched and annealed randomness of interaction graph (see legend), brought about by the sufficiently large fraction $\nu$ of nodes that have a lower site-specific invasion rate $w_j=0.1$ than the rest of the population at $w_j=1$. Presented results demonstrate that quenched site-specific invasion rates drawn from a simple discrete double-peaked distribution are apt to fully suppress global oscillations (note that the value of the order parameter $A$ drops to zero at a sufficiently large value of $\nu$ in both cases), regardless of the type of randomness in the interaction network that enables them. We have used $Q=0.5$ for the fraction of rewired links determining the level of quenched randomness, and $P=0.5$ as the probability for a potential target of an invasion to be selected randomly from the whole population rather than from nearest neighbors determining the level of annealed randomness of network.}
\end{figure}

In Fig.~\ref{nu}, we present representative results for both quenched and annealed randomness of interaction graph (see legend). The first observation is that only a minute fraction of suppressed nodes (less than $5 \%$) suffices to fully suppress global oscillations, and this regardless of the applied high $Q$ and $P$ values that practically ensure an optimal support for local oscillations to synchronize across the population into global oscillations. Moreover, it can be observed that both transitions to the oscillation-free state are continuous. In other words, there does not exist a sharp drop in the value of $A$ at a particular value of $\nu$. Instead, the suppression of global oscillations is gradual as the level of site-specific invasion heterogeneity in the population increases.

\begin{figure}
\centerline{\epsfig{file=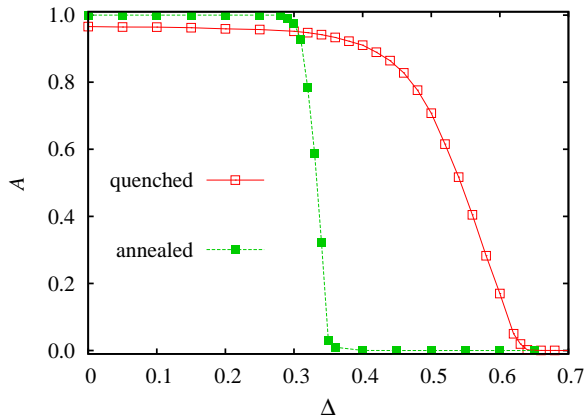,width=8.5cm}}
\caption{\label{dif} The suppression of global oscillations in the presence of quenched and annealed randomness of interaction graph (see legend), brought about by the sufficiently large difference $\Delta$ between the quenched site-specific invasion rates in the double-peaked discrete distribution. Here a fraction $\nu=0.1$ of the population has a lower site-specific invasion rate equal to $w_j=1-\Delta$, while the remainder of the population has $w_j=1$. Similarly as in Fig.~\ref{nu}, presented results demonstrate that quenched site-specific invasion rates fully suppress global oscillations (the value of $A$ drops to zero at a sufficiently large value of $\Delta$ in both cases), regardless of the type of randomness in the interaction network that enables them. Likewise as in Fig.~\ref{nu}, we have used $Q=0.5$ for the level of quenched randomness and $P=0.5$ for the level of annealed randomness of topology.}
\end{figure}

Similar in spirit, another way to introduce invasion heterogeneity gradually into the population is to use a fixed fraction of nodes with a lower invasion rate, but vary the difference to $w_j=1$. Accordingly, we have a fraction $\nu=0.1$ of nodes, which instead of $w_j=1$ have the invasion rate $w_j=1-\Delta$. Here $\Delta$ becomes the free parameter, which for zero returns the traditional rock-paper-scissors game with homogeneous invasion rates, while for $\Delta>0$ the distance in the peaks of the discrete double-peaked distribution, and thus the level of heterogeneity in the population, increases. Representative results obtained with this approach are shown in Fig.~\ref{dif} for both quenched and annealed randomness of interaction graph (see legend). In comparison with results presented in Fig.~\ref{nu}, it can be observed that increasing $\Delta$ has somewhat different consequences than increasing $\nu$. In the former case, when $\Delta$ is small, the slight heterogeneity has no particular influence on the stationary state and global oscillations persist well beyond $\Delta=0.3$ for annealed randomness and $\Delta=0.6$ for quenched randomness. But if the difference reaches a sufficiently large value, global oscillations disappear in much the same gradual way as observed before in Fig.~\ref{nu}, although the transition for annealed randomness is more sudden.

To sum up our observations thus far, different versions of the same concept reveal that spatially quenched heterogeneity in site-specific invasion rates is capable to effectively suppress global oscillations that would otherwise be brought about by either annealed or quenched randomness in the interaction network. However, there is yet another possible source or large-amplitude global oscillations in the population, namely mobility. As is well-known, mobility can give rise to global oscillation that jeopardizes biodiversity \cite{reichenbach_n07}. Although subsequent research revealed that global oscillations due to mobility do not emerge if the total number of competing players is conserved \cite{peltomaki_pre08}, more recently it was shown that, if in addition to a conservation law also either quenched or annealed randomness is present in the interaction network, then mobility still induces global oscillations \cite{szolnoki_pre16}. In particular, if the site exchange is intensive then only a tiny level of randomness in the host lattice suffices to evoke global oscillations.

\begin{figure}
\centerline{\epsfig{file=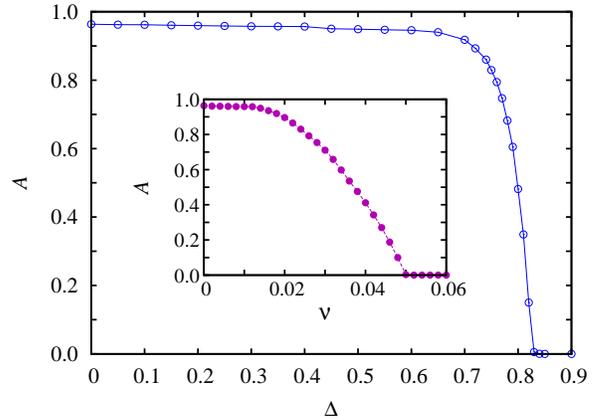,width=8.5cm}}
\caption{\label{mob} The suppression of global oscillations in the presence of mobility, brought about by the sufficiently large difference $\Delta$ between the quenched site-specific invasion rates in the double-peaked discrete distribution, and by the sufficiently large fraction $\nu$ of nodes that have a lower site-specific invasion rate $w_j=0.1$ than the rest of the population at $w_j=1$ (inset). In the main panel, a fraction $\nu=0.1$ of the population has $w_j=1-\Delta$, while the remaining $90 \%$ have $w_j=1$. Moreover, a rather strong rate of mobility at $\sigma=0.9$, along with a $Q=0.05$ level of quenched randomness to ensure global oscillations in the absence of heterogeneity in site-specific invasion rates is used in both the main panel and the inset. Evidently, although slightly higher thresholds in $\nu$ and $\Delta$ then recorder in Figs.~\ref{nu} and \ref{dif} are needed, in both cases quenched heterogenous site-specific invasion rates fully suppress global oscillations as the value of $A$ eventually drops to zero in the main panel and the inset.}
\end{figure}

Lastly, we thus verify if heterogeneity in site-specific invasion rates is able to suppress global  oscillations brought about by mobility. As results in Fig.~\ref{mob} show, the impact of quenched invasion heterogeneity is very similar to the above-discussed cases. It is worth noting that conceptually similar behavior can be observed when biological species are hosted in a turbulent flow of fluid environment \cite{karolyi_jtb05, groselj_pre15}. In fact, as a general conclusion, neither randomness in the interaction network nor the mobility of players can compensate for the detrimental impact of spatial invasion heterogeneity on global oscillations, thus establishing the latter as a very potent   proponent of biodiversity in models of cyclic dominance.

\section*{Discussion}
We have studied the impact of site-specific heterogeneous invasion rates on the emergence of global oscillations in the spatial rock-paper-scissors game. We have first confirmed that species-specific heterogeneous invasion rates, either fixed or varying over time, fail to disrupt the synchronization of locally emerging oscillations into a global oscillatory state on a regular small-world network. On the contrary, we have then demonstrated that site-specific heterogeneous invasion rates, determined once at the start of the game, successfully hinder the emergence of global oscillations and thus preserves biodiversity. We have shown this conclusion to be valid independently of the properties of the distribution that determines the invasion heterogeneity, specifically demonstrating the failure of coordination for uniformly and double-peak distributed site-specific invasion rates. Moreover, our research has revealed that quenched site-specific heterogeneous invasion rates preserve biodiversity regardless of the type of randomness that would be responsible for the emerge of global oscillations. In particular, we have considered quenched and annealed randomness in the interaction network, as well as mobility. Regardless of the type of randomness that would promote local oscillations to synchronize across the population to form global oscillations, site-specific heterogeneous invasion rates were always found to be extremely effective in suppressing the emergence of global oscillations. Drawing from the colloquial expression used to refer to alcohol that is consumed with the aim of lessening the effects of previous alcohol consumption, the introduction of randomness in the form of site-specific heterogeneous invasion rates lessens, in fact fully suppresses, the effects of other types of randomness -- hence the ``hair of the dog'' phenomenon.

Our setup takes into account heterogeneities that are inherently present in virtually all uncontrolled environments, ranging from bacterial films to plant communities. Examples include qualitative and quantitative variations in the availability of nutrients, local differences in the habitat, or any other factors that are likely to influence the local success rate of the governing dynamics \cite{elowitz_n00, cameron_jecol09}. The consideration of site-specific heterogeneous invasion rates and their ability to suppress global oscillations joins the line of recent research on the subject, showing for example that the preservation of biodiversity is promoted if a conservation law is in place for the total number of competing players \cite{peltomaki_pre08}, or if zealots are introduced to the population \cite{verma_pre15, szolnoki_pre16}. Notably, previously it was shown that zealotry can have a significant impact on the segregation in a two-state voter model \cite{mobilia_prl03}, and research in the realm of the rock-paper-scissors game confirmed such an important role of this rather special uncompromising behavior. In general, since global, population-wide oscillations are rarely observed in nature, it is of significance to determine key mechanisms that may explain this, especially since factors that promote such oscillations, like small-world properties, long-range interactions, or mobility, are very common. In this sense, site-specific heterogeneous invasion rates fill an important gap in our understanding of the missing ingredient that would prevent local oscillations to synchronize across the population to form global oscillations.

There is certainly no perfect spatial system where microscopic processes would unfold identically across the whole population. These imperfections are elegantly modeled by the heterogeneous host matrix that stores the individual invasion rates of each player. As we have shown, the coordination of species evolution is highly sensitive on such kind of heterogeneities when they are fixed in space. Ultimately, this prevents the synchronization of locally emerging oscillations, and gives rise to a ``hair of the dog''-like phenomenon, where one type of randomness is used to mitigate the adverse effects of other types of randomness. We hope that these theoretical explorations will help us to better understand the rare emergence of global oscillations in nature, as well as inspire further research, both experimental and theoretical, along similar lines.

\section*{Methods}
The spatial rock-paper-scissors game evolves on a $L \times L$ square lattice with periodic boundary conditions, where each site $j$ is initially randomly populated by one of the three competing species. For convenience, we introduce the $S_i \to S_{i+1}$ notation, where $i$ runs from $0$ to $2$ in a cyclic manner. Hence, species $S_0$ (for example paper) invades species $S_1$ (rock), while species $S_1$ invades species $S_2$ (scissors), which in turn invades species $S_0$ to close the loop of dominance. The evolution of species proceeds in agreement with a random sequential update, where during a full Monte Carlo step ($MCS$) we have chosen every site once on average and a neighbor randomly. In case of different players the invasion was executed according to the rock-scissors-paper rule with probability $\lambda$.

In the simplest, traditional version of the game, all invasion rates between species are equal to $\lambda=1$. Species-specific heterogeneous invasion rates can be introduced through the parameter $\lambda=\delta_i$, which is simply the probability for the $S_i \to S_{i+1}$ invasion to occur when given a chance. The values $\delta_0$, $\delta_1$ and $\delta_2$ can be determined once at the start of the game, or they can be chosen uniformly at random from the unit interval at each particular instance of the game. On the other hand, site-specific heterogeneous invasion rates, which we denote as $w_j$, apply to each site $j$ in particular, and determine the probability that a neighbor will be successful when trying to invade player $j$ according to the $S_i \to S_{i+1}$ rule. This rule can be considered as "prey-dependent" because the $w$ value at the prey's position determines the probability of invasion. As an alternative rule, we can consider the $w$ value of predator's position that determines the invasion probability. Lastly, we can assume that the $w$ values of both the predator and prey's positions influence the invasion rate via their $w_j \cdot w_{j\prime}$ product. While the time dependence of the evolution will be different in the mentioned three cases but the qualitative behavior is robust. Therefore we restrict ourself to the first mentioned "prey-dependent" rule.

These invasion rates are determined once at the start of the game and can be drawn uniformly at random from the unit interval, or from any other distribution. Here, in addition to uniformly distributed $w_j$, we also consider site-specific heterogeneous invasion rates drawn from a discrete double-peaked distribution, where a fraction $\nu$ of sites have $w_j=1-\Delta$, while the remaining $1-\nu$ have $w_j=1$.

To test the impact of site-specific heterogeneous invasion rates under different circumstances, we consider interaction randomness in the form of both quenched and annealed randomness. Quenched randomness is introduced by randomly rewiring a fraction $Q$ of the links that form the square lattice whilst preserving the degree $z=4$ of each site. This is done only once at the start of the game. This procedure returns regular small-world networks for small values of $Q$ and a regular random network in the $Q \to 1$ limit \cite{szabo_jpa04}. Annealed randomness, on the other hand, is introduced so that at each instance of the game a potential target for an invasion is selected randomly from the whole population with probability $P$, while with probability $1-P$ the invasion is restricted to a randomly selected nearest neighbor \cite{szolnoki_pre04b}. This procedure returns well-mixed conditions for $P=1$, while for $P=0$ only short-range invasions as allowed by the original square lattice are possible.

We also consider the impact of site-specific heterogeneous invasion rates in the presence of mobility. The latter is implemented so that during each instance of the game we choose a nearest-neighbor pair randomly where players exchange their positions with probability $\sigma$. Oppositely, with probability $1-\sigma$, the dominant species in the pair invades the other in agreement with the rules or the rock-paper-scissors game. The parameter $\sigma$ hence determines the intensity of mobility while the number of players is conserved. Technically, however, the strategy exchange between neighboring players is determined not only by the level of mobility $\sigma$, but it also depends on the individual $w_j$ and $w_r$ values characterizing the neighboring sites $j$ and $r$. In this way, we can consider the fact that different sites may be differently sensitive to the change of strategy, and the success of mutual change is then practically determined by the site that is more reluctant to change its state. Accordingly, when the strategy exchange is supposed to be executed, then this happens only with the probability $w_{jr}$ that is equal to the smaller of $w_j$ and $w_r$ values (all the other details of the model remain the same as above).

Global oscillations are characterized with the order parameter $A$, which is defined as the area of the limit cycle in the ternary diagram \cite{szabo_jpa04}. This order parameter is zero when each species occupies one third of the population, and becomes one when the system terminates into an absorbing, single-species state. We have used lattices with up to $L \times L = 4 \cdot 10^6$ sites, which was large enough to avoid accidental fixations when the amplitude of oscillations was large, and which allowed an accurate determination of strategy concentrations that are valid in the large population size limit. Naturally, the relaxation time depends sensitively on the model parameters and the system size, but $10^5$ MCS was long enough even for the slowest evolution that we have encountered during this study.

\end{document}